\documentclass[twocolumn,showpacs,preprintnumbers,amsmath,amssymb]{revtex4}
\usepackage{amsfonts}
\usepackage{amsmath}
\usepackage{graphicx}%Include figure files
\usepackage{dcolumn}%Align table columns on decimal point
\usepackage{bm}
\usepackage{overpic}
\usepackage{booktabs}
\usepackage{color}

\begin{document}
\title{Spectral coarse graining for random walk in bipartite networks}
\author{Yang Wang$^{1,2}$, An Zeng$^{3}$\footnote{an.zeng@unifr.ch}, Zengru Di$^{1,2}$ and Ying Fan$^{1,2}$\footnote{yfan@bnu.edu.cn}}

\affiliation{1 Department of Systems Science, School of Management, Beijing Normal University, Beijing 100875, P.R. China\\
2 Center for Complexity Research, Beijing Normal University, Beijing 100875, P.R. China\\
3 Department of Physics, University of Fribourg, Chemin du Mus\'{e}e 3, CH-1700 Fribourg, Switzerland}

\date{\today}
\begin{abstract}
Many real-world networks display a natural bipartite structure, while
analyzing or visualizing large bipartite networks is one of the most
challenges. As a result, it is necessary to reduce the complexity
of large bipartite systems and preserve the functionality at the same time. We observe, however, the existing coarse graining methods for binary networks fail to work in the bipartite networks. In this paper, we use the spectral analysis to design a coarse graining scheme specifically for bipartite networks and keep their random walk properties unchanged. Numerical analysis on artificial and real-world bipartite networks indicates that our coarse graining scheme could obtain much smaller networks from large ones, keeping most of the relevant spectral properties. Finally, we further validate the coarse graining method by directly comparing the mean first passage time between the original network and the reduced one.
\end{abstract}

\maketitle
\section{Introduction}
%% introduction of complex networks
As a backbone of many complex systems, complex networks have been intensively studied in the past decade. Examples range from social relationships among individuals, to interactions of proteins in biological systems, to the interdependence of function calls in large software projects. The network analysis has greatly helped us understand the structure and function of real-world systems~\cite{Rev.Mod.Phys.74,SIAM Rev.45,Science 286,Nature 393,Phys. Rep. 424,Newman2010}.

%%bipartite networks
Bipartite network is an important kind of complex network, which is composed of two types of nodes with no links connecting nodes of the same type. For example, the e-commercial systems consisting of online users and products~\cite{EPL9718005,PhysicaA3911822}, the scientific collaboration system consisting of authors and papers~\cite{Newman2001(1),Newman2001(2)}, and family name inheritance system consisting of babies and names~\cite{PRE75056101} are naturally described by such networks. Recently, some topological properties such as clustering coefficient and modularity of bipartite networks have been studied~\cite{PRE72056127,Zhang2007,Kitsak2011}. However, one of the most difficult hurdles in analyzing and visualizing bipartite network is the size of the real-world systems. The online commercial systems, for instance, can have thousands of products and even millions of users. Given that most of the algorithms which are used to extract the properties of the bipartite network run in times that grow polynomially with the system size $N$, systems with huge sizes become a challenge.

%%reviewing coarse graining methods
In order to deal with the problem mentioned above, a promising way is to consider some units of the system as almost indistinguishable and to merge them into one, i.e., to reduce the number of nodes and edges by means of a mapping of the network with $N$ nodes and $E$ edges into a smaller one with $\bar N$ nodes and $\bar E$ edges. Based on this concept, several coarse graining schemes for binary networks including $k$-core decomposition~\cite{Bolloba1984,NP6888}, box-covering process~\cite{Song2005,Goh2006}, geographical coarse graining~\cite{PRL93168701}, spectral coarse graining~\cite{Gfeller2008,Gfeller2007} have been proposed.

Specifically, the $k$-core decomposition intends to classify nodes into different shells which represent their importance. This technique can be used to isolate the central core of a network, and was also shown to be extremely effective for visualization purposes. The box-covering technique yields a new network which can preserve some of the topological features of the original ones. The geographical coarse graining uses a renormalization-group like numerical analysis to reduce the size of the networks while preserving the degree distribution, clustering coefficient and assortativity correlation. The spectral coarse graining methods, on the other hand, focus on the dynamic processes taking place on networks. They merge nodes based on the eigenvectors of different matrices, so that some dynamic properties such as random walk and synchronization of the original network are kept unchanged. Mathematically, the spectral-based methods are expressed as preserving some eigenvalues of the stochastic matrix or the graph Laplacian. Furthermore, some works have been further dedicated to coarse grain networks for dynamics of heterogeneous oscillators~\cite{PRE84036708} and other critical phenomena~\cite{PRE82011107}.

Actually, a very close related problem is the community detection which groups nodes based on link density. Because of the importance and the complexity of finding meaningful communities, it is a fact that recent years have witnessed an explosion of research on community structure in graphs, and a huge number of methods or techniques have been designed~\cite{PNAS99,Phys. Rev. E 74,Phys.Rev.E 72,Proc. Natl. Acad. 103,PRE82066106,Rhys. Rev. E 77,PRE81046110} (see~\cite{Physics Reports486} for a review). However, there is often no clear statement on which properties of the initial network are preserved in the network of clusters.

%%the limitation of previous method
Though the coarse graining methods mentioned above are effective in binary networks, they usually have some limitations when extended to directed and bipartite networks. In directed networks, the role of nodes in dynamics cannot be well characterized by the eigenvectors since imaginary value emerges when the adjacency and Laplacian matrix are asymmetric. This problem can be solved by directly using the paths to determine the similarity between nodes and finally preserve the dynamics properties (synchronization) when merging nodes~\cite{PRE83056123}. For bipartite networks, the situation can be even more complicated. There are two types of nodes in bipartite networks and the dynamics on both types of nodes should be preserved. More importantly, the coarse graining method should preserve the intrinsic bipartite feature of the networks (i.e. no link exists between nodes in each same type). However, if we regard the bipartite networks as binary ones and apply previous coarse graining methods, it will end up with merging nodes from different types into one. Furthermore, using the existing community detection methods to coarse grain bipartite networks may significantly change the network function~\cite{Zhang2007,Lehmann2008}. As a result, it is still a challenge to preserve both the network function and the bipartite property.

%%what we do
In this paper, employing random walks to be the main characteristic~\cite{Noh2004}, we introduce a new spectral-based approach to coarse grain bipartite networks. Unlike the coarse graining methods for binary networks, our goal is to obtain a reduced bipartite network that preserves both the original random walk properties and the bipartite feature. In order to preserve the random walk properties of both types of nodes, two matrices (denoted by $\textbf{W}_m$ and $\textbf{W}_n$) based on the stochastic matrix of the bipartite network are introduced, and a new coarse grain scheme relied on $\textbf{W}_m$ and $\textbf{W}_n$ is designed. The obtained network remains bipartite and have very similar spectral properties to the original bipartite network. Moreover, we validate our method by performing a direct test of the mean first passage time (MFPT) to artificial and real-world bipartite networks. The new method is robust in various kinds of bipartite networks and the choices of sinks. Finally, we remark that this method can be easily extend to preserve many other of spectral-determined dynamical properties in bipartite networks.

\section{Spectral coarse graining method on bipartite networks}

\subsection{random walks on binary networks}
Random walks play a central role in dynamical properties taking place on complex networks. Starting at some specified initial vertices, the walker jumps with equal probability from its current location to one of its nearest neighbors at each time step. Considering a binary network $G=(V,E)$ with $N$ nodes. The adjacency matrix $\textbf{A}$ is the matrix with elements $A_{ij}=1$ if there is an edge connecting vertices $i$ and $j$, otherwise $0$. Let $p_i(t)$ be the probability that the walker is at vertex $i$ at time step $t$. If the walker is at vertex $j$ at time step $t-1$, the probability of taking a jump along any of its neighbors is $1/k_j$. Accordingly, $p_i(t)$ on an undirected binary network is given by
\begin{equation}
\label{equation1}
p_i(t) = \sum_j \frac{A_{ij}}{k_j}p_j(t - 1),
\end{equation}
where $k_j$ is the degree of vertex $j$. As a matrix form, Eq. \ref{equation1} can be written as $\vec{p}(t) = \textbf{A}\textbf{D}^{-1} \vec{p}(t - 1)$ where $\vec{p}$ is the vector with elements $p_i$ and \textbf{D} is the diagonal matrix with the degrees of the vertices down its diagonal $\textbf{D} = diag(d_1,d_2,...,d_N)$. Defining a stochastic matrix $\textbf{W} = \textbf{D}^{-1}\textbf{A}$, random walk in binary network can be characterized by the stochastic matrix \textbf{W}, and the element $w_{ij}$ describes the probability that a walker starts from node $i$ to node $j$.

In the context of transport phenomena or search on a network, MFPT (mean first passage time) is an important characteristic of random walks~\cite{Baronchelli2006,Noh2004}. To compute it exactly, one usually considers some nodes as traps. The normalized Laplacian matrix of the network is defined as $\textbf{L} = \textbf{I} - \textbf{D} ^ {-1} \textbf{A}$, where \textbf{I} is the identity matrix. We use $\Gamma$ to denote the set of traps and $|\Gamma|$ to represent the number of traps. For simplicity, we distinguish all nodes in the network by assigning each of them a unique number. We label consecutively all nodes, excluding those in $\Gamma$, from 1 to $N-\Gamma$ and sinks are labeled from $N-\Gamma+1$ to $N$. By suppressing the last $|\Gamma|$ rows and columns of the normalized Laplacian matrix, we obtain a submatrix of the normalized Laplacian matrix \textbf{L} as $\textbf{L}'$.

It is shown in~\cite{Zhang2012} that the trapping time or the first passage time $T_i$, which is defined as a particle first arriving at any one of the traps, given that it
starts from node $i$, can be expressed as follows,
\begin{equation}
\label{equation2}
T_i = \sum_{j = 1}^{N - |\Gamma|} {l_{ij}^{ - 1}},
\end{equation}
where $l_{ij} ^ {-1} $ is the elements of matrix $\textbf{L}'$. Then the MFPT $\langle T\rangle$, which
is defined as the average of $T_i$ over all initial nodes distributed uniformly over nodes including the traps, is given by
\begin{equation}
\label{equation3}
\langle T\rangle = \frac{1}{N} \sum_{i = 1}^{N - |\Gamma|} T_i = \frac{1}{N} \sum_{i = 1}^{N - |\Gamma|} \sum_{j = 1}^{N - |\Gamma|} {l_{ij}^{ - 1}}.
\end{equation}
Eq. \ref{equation3} can also be found in the literature in several equivalent forms\cite{Kemeny1976,Aldous_web}.

Thus the exact solution of the MFPT of the unbiased random walk is given, independently from the number and the location of the sinks. The Eq.~\ref{equation2} and Eq.~\ref{equation3} are very necessary since they can reduce the problem of computing the MFPT to calculating the inverse matrix $\textbf{L}'$ and also can be used to check the MFPT of different networks in the following sections, at least for networks with relatively small size.

\subsection{random walks on bipartite networks}

In bipartite networks, connection between vertices is also described by the adjacency matrix. However, since a bipartite network consists of two non-overlapping kinds of nodes and the links can only exist between two nodes from distinct sets. The adjacency matrix $\textbf{A}$ of a bipartite network is defined as a matrix with order $M \times N$, where $M$ and $N$ are the number of vertices of these two distinct sets. In this paper, we call two types of nodes as top and bottom nodes, respectively. If there is a link between vertices $i$ in the top sets and $j$ in the bottom sets, the element $A_{ij} = 1$, otherwise $A_{ij} = 0$. In bipartite networks, the random walk process is closely related to the information filtering algorithms~\cite{PNAS1074511,PRE83066119}. Unlike binary networks, the stochastic matrix of a bipartite network is divided into two matrices. If a walker goes from the top set to the bottom set, the stochastic matrix $\textbf{U}$ is with order $M \times N$ with elements $U_{ij} = A_{ij}/k_i$ which describe the probability from node $i$ in the top set to node $j$ in the bottom set. If the walker is from the bottom set to the top set, then $\textbf{V}$ is with order $N \times M$ in which $V_{ij}=A_{ji}/k_i$. $\textbf{U}$ and $\textbf{V}$ contained all the information of random walk in bipartite network.

Now we define two new matrices $\textbf{W}_m$ and $\textbf{W}_n$ as follows: $\textbf{W}_m = \textbf{U} \times \textbf{V}$ and $\textbf{W}_n = \textbf{V} \times \textbf{U}$. Just like the stochastic matrix in binary networks, $\textbf{W}_m$ and $\textbf{W}_n$ are square matrices. $\textbf{W}_m$ ($\textbf{W}_n$) describes the random walkers going from top (bottom) nodes to top (bottom) nodes. These two matrices have some interesting properties. In particular, the largest eigenvalue of these two matrices is equal to $1$ and the corresponding eigenvector is constant. Moreover, there are several largest eigenvalues of these two matrices with the same value. As discussed in~\cite{Gfeller2007}, eigenvectors corresponding to the eigenvalues close to $1$ of the stochastic matrix \textbf{W} capture the large-scale behavior of the random walk in binary network. The fact is also true in $\textbf{W}_m$ and $\textbf{W}_n$ in bipartite network since they are square matrices just like \textbf{W}. As a result, our goal is to preserve the largest eigenvalues and eigenvectors of $\textbf{W}_m$ and $\textbf{W}_n$. In this way, we can preserve the properties of random walk in bipartite network.

\subsection{Spectral coarse graining method for bipartite networks}

First of all, two nodes $i$ and $j$ with exactly the same neighbors should be merged since they cannot be distinguished from the point of view of random walk. In terms of a eigenvector $\vec{p}_{\alpha}$ for any $\lambda_{\alpha} \neq 0$ of $\textbf{W}_m$ or $\textbf{W}_n$ which is implied that $p_{\alpha}^i = p_{\alpha}^j$. Here, we denote the eigenvalues of a matrix $\textbf{W}_m$ or $\textbf{W}_n$ as $\lambda_{\alpha}$ and their corresponding eigenvectors $\vec{p}_{\alpha}$. After merging, the new node will carry the sum of the edges of nodes $i$ and $j$ and the resulting adjacency matrix of a bipartite network $\tilde{A}$ will have order $(M - 1) \times N$ or $M \times (N - 1)$, with the corresponding line or column of the new node being the sum of the line (column) $i$ and $j$. The properties of random walk in the new bipartite network is exactly the same as those in the original network. Moreover, if $p_{\alpha}^i \approx p_{\alpha}^j$ we could also group them in order to obtain an even smaller bipartite network. By definition, if $|p_{\alpha}^i - p_{\alpha}^j| \propto \epsilon$ we could group node $i$ and $j$ together. Like ref.~\cite{Gfeller2007,Gfeller2008}, the condition $|p_{\alpha}^i - p_{\alpha}^j| \propto \epsilon$ can be implemented by defining a parameter $I$ as equally distributed intervals between the minimize and the maximum components of each eigenvector $\vec{p}$. The nodes whose eigenvector components in $\vec{p}$ fall in the same interval should be grouped.

We summarize the bipartite network spectral coarse graining (BSCG) method in the following procedures:

\begin{enumerate}
\item For any given bipartite network \textbf{A}, we can get two stochastic matrices $\textbf{U}$ and $\textbf{V}$ which gives the transition probability from the top nodes to bottom nodes and bottoms nodes to top nodes in bipartite network, respectively;
\item Using $\textbf{U}$ and $\textbf{V}$, we can obtain two square stochastic matrices $\textbf{W}_m = \textbf{U} \times \textbf{V}$ and $\textbf{W}_n = \textbf{V} \times \textbf{U}$.
\item Calculating the eigenvalues $\lambda_{\alpha}$ and the corresponding eigenvectors $\vec{p}_{\alpha}$ of $\textbf{W}_m$ and $\textbf{W}_n$;
\item Merging nodes with similar components in the $\vec{p}_{\alpha}$ as one node. In the new adjacency matrix $\tilde{A}$, this node will carry the sum of the edges of original nodes. The nodes in the top set should be merged based on the eigenvectors of $\textbf{W}_m$ and the nodes in the bottom set should be merged based on the eigenvectors of $\textbf{W}_n$.
\end{enumerate}

The obtained adjacency matrix $\tilde{\textbf{A}}$ is a weighted matrix. The new stochastic matrices $\widetilde{\textbf{U}}$ and $\widetilde{\textbf{V}}$ can be calculated as $\widetilde{U}_{ij}= \tilde{A}_{ij}/\sum_j{\tilde{A}_{ij}}$ and $\widetilde{V}_{ij}= \tilde{A}_{ji}/\sum_j{\tilde{A}_{ji}}$. This method can be further extended to more than one eigenvector. In this case, groups are defined as nodes with almost the same component over the eigenvectors corresponding to the largest nontrivial eigenvalues. Actually, choosing several largest nontrivial eigenvalues could better preserve the properties of random walk in bipartite network.

\section{Results}

To validate the new method, we apply it to both artificial and real-world bipartite networks.

\subsection{Artificial Networks}

To begin with, let us consider an artificial bipartite networks with 1200 vertices, which are divided into 2 sets. The top set has 300 vertices and the bottom set has 900 vertices, and nodes in each set are divided into 10 groups with equal size. In other words, each group in the top set has 30 vertices while there are 90 vertices in each group of the bottom set. The probability for existing a link between each pair of nodes in the same group is $p_1$ while $p_2$ is the corresponding probability outside the group. In this section, $p_1 = 0.4$ and $p_2 = 0.05$. Since this kind of artificial networks have significant community structure, we call them community networks. Using the clustering method in original bipartite network~\cite{Zhang2007}, 10 communities are correctly detected from the network.

To coarse grain this bipartite network, we used the largest three nontrivial eigenvectors $\vec{p}_2$, $\vec{p}_3$ and $\vec{p}_4$. We set $I = 12$, which means that 12 intervals with equal size are divided between the largest and the lowest component of each eigenvector. Using the BSCG method, we get a rather small network with 391 vertices. Clearly, since the BSCG method and community detection method focus on different properties in the bipartite network, the nodes grouping results are different. Table~\ref{Table1} shows the three largest nontrivial eigenvalues of $\textbf{W}_m$ and $\textbf{W}_n$ before and after the coarse graining. Though the largest nontrivial eigenvalues of these two matrices are the same, note that the matrices $\textbf{W}_m$ and $\textbf{W}_n$ contain different information. Specifically, $\textbf{W}_m$ is with order $M \times M$ and describes the probability that the walker is from the top set to the top set, while $\textbf{W}_n$ considers the information that the walker is from the bottom set to the bottom set. Moreover, the eigenvectors of the largest nontrivial eigenvalues of $\textbf{W}_m$ and $\textbf{W}_n$ are respectively corresponding to the nodes in the top set and bottom set. Thus, the eigenvectors of both matrices should be considered and the nontrivial eigenvalues of both matrices should be preserved. The eigenvalues of $\widetilde{\textbf{W}}_m$ and $\widetilde{\textbf{W}}_n$ of the reduced network can be calculated from the reduced adjacency matrix $\tilde{\textbf{A}}$, using the same way described in the previous section. As expected from our perturbative derivation, the largest three eigenvalues are effectively preserved in the coarse-grained network.

Moreover, we also apply the BSCG method to ER random bipartite networks and obtain similar results (see also Table~\ref{Table1}). In the ER random bipartite networks, the probability of having a link between two vertices of different sets is $p = 0.01$ and the top set contains 1000 vertices while bottom set has 800 vertices. In the ER bipartite network, we also focus on the three largest nontrivial eigenvectors and set $I=20$. The results in Table~\ref{Table1} indicates that our new method is robust in various kinds of artificial networks.

\begin{table}[htbp]
\begin{center}

\caption{The three largest nontrivial eigenvalues of $\textbf{W}_m$ and $\textbf{W}_n$ in the artificial including Community bipartite network and ER random bipartite network. $\lambda_{\alpha}$ and $\tilde{\lambda_{\alpha}}$ are the eigenvalues before and after coarse graining, respectively.}
\label{Table1}
\begin{tabular}{c c c c c c}
\hline
\hline
Network & $\alpha$ & $\lambda_{\alpha}$ ($\textbf{W}_m$) & $\tilde{\lambda_{\alpha}}$ ($\textbf{W}_m$) & $\lambda_{\alpha}$ ($\textbf{W}_n$) & $\tilde{\lambda_{\alpha}}$ ($\textbf{W}_n$)  \\
\hline
 & 2  & 0.4405& 0.4336 & 0.4405& 0.4336\\
Community network & 3  &0.4342 & 0.4279 & 0.4342 & 0.4279\\
 & 4 & 0.4180 & 0.4076 &  0.4180 & 0.4076\\
 \hline
  & 2  & 0.3986& 0.3933 & 0.3986& 0.3933\\
ER network & 3  &0.3908 & 0.3833 & 0.3908 & 0.3833\\
 & 4 & 0.3865 & 0.3784 &  0.3865 & 0.3784\\
\hline
\hline
\end{tabular}
\vspace*{0.0cm}
\end{center}
\end{table}

\subsection{Real-world Bipartite Networks}

\begin{figure}[htb]
  \begin{minipage}[b]{\columnwidth}
    \centering
    \includegraphics[width=\columnwidth]{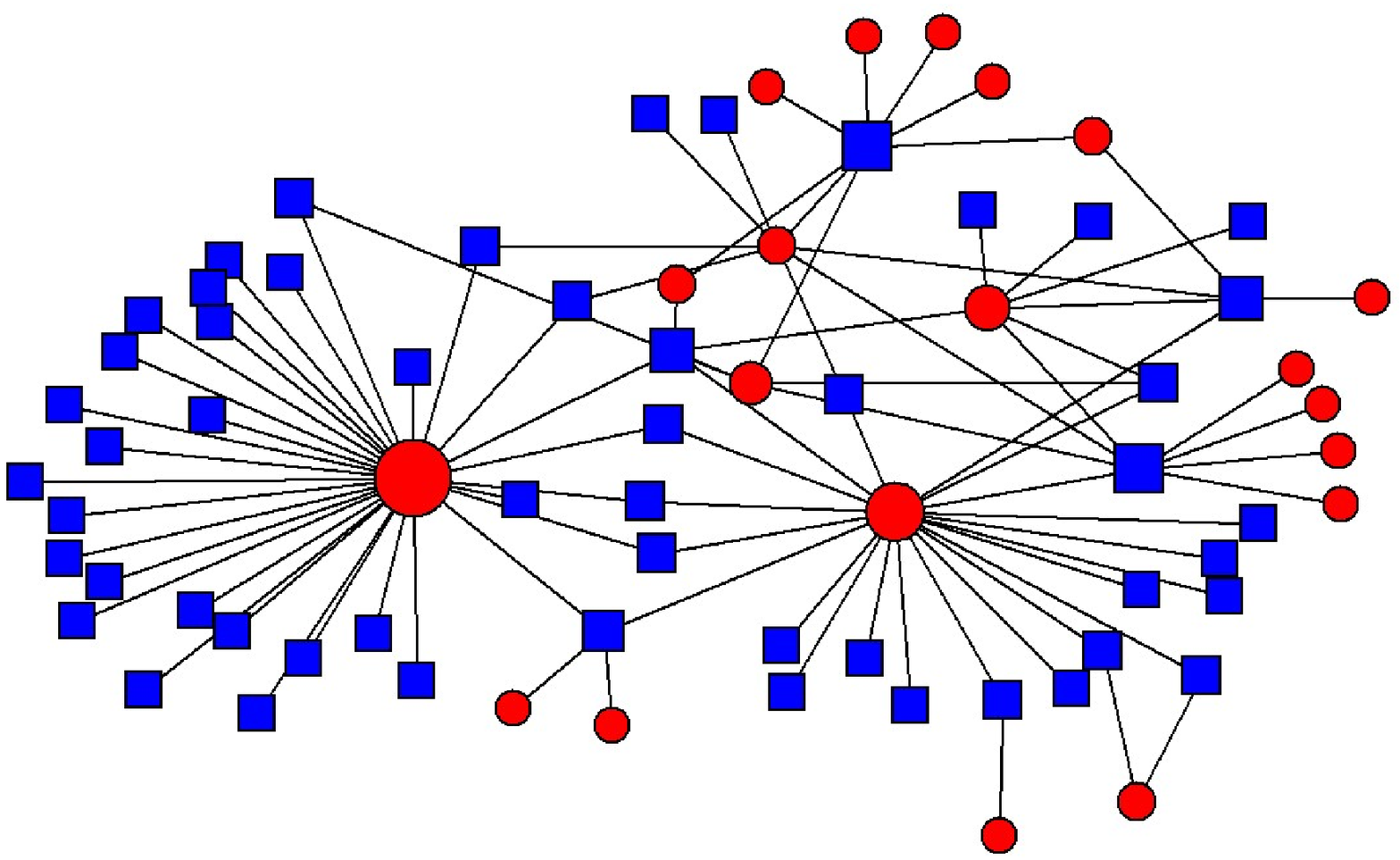}
  \end{minipage}\\
    \begin{minipage}[b]{\columnwidth}
    \centering
    \includegraphics[width=\columnwidth]{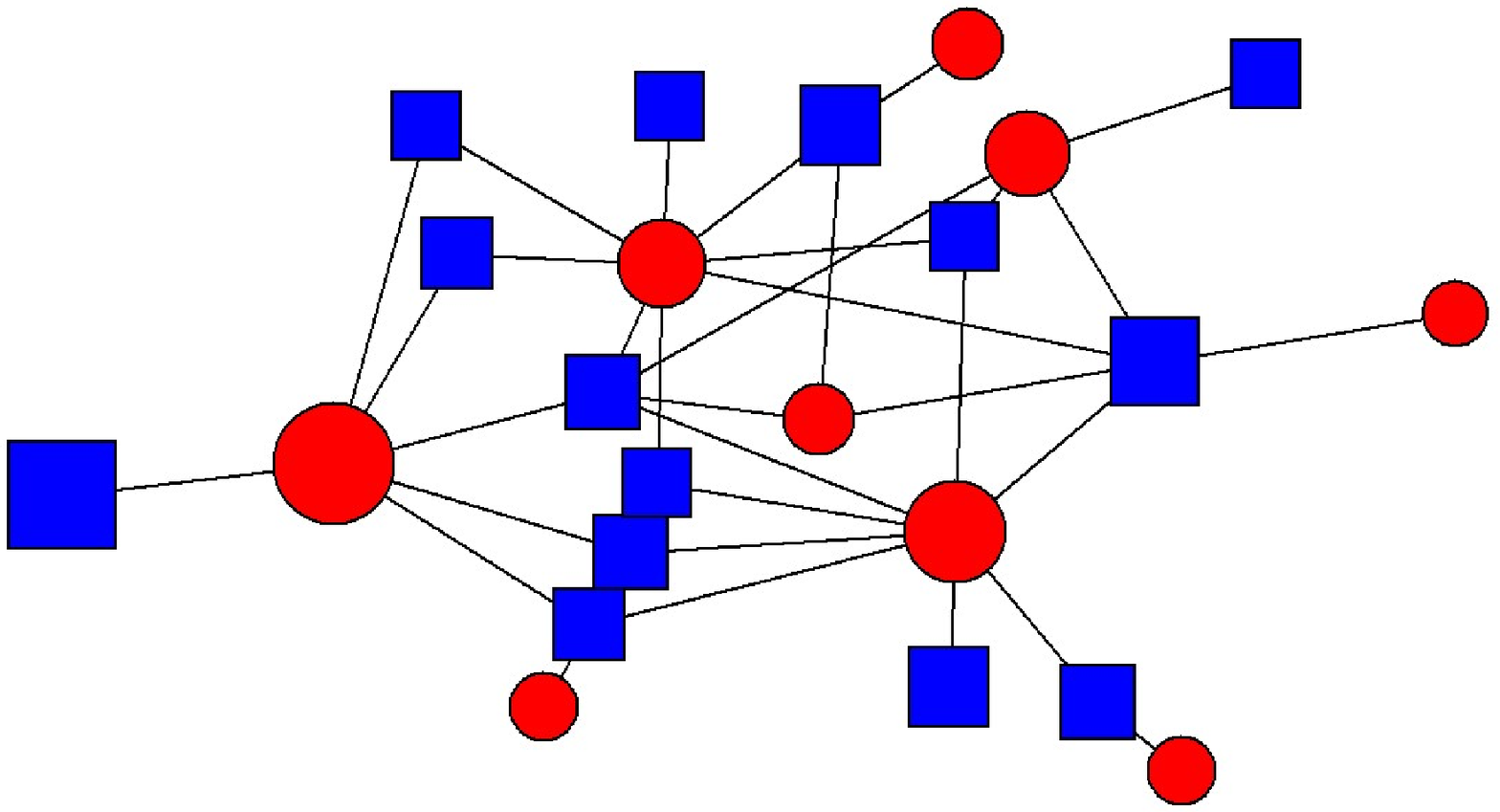}
  \end{minipage}
  \caption{(Color online) The top figure is a social bipartite networks of terrorists with $N = 73$. Node size is proportional to their degree. The two different colors represent two kinds of vertices. The blue one indicates the people and the red one represents the organizations they belong to.
The Bottom figure is the coarse-grained network ($N = 23$) according to our method based on random walk properties. Node size is proportional to the its strength in this weighted network. Colors still correspond to these two kinds of vertices.}
\end{figure}

\begin{table}[htbp]
\begin{center}
\caption{The three largest nontrivial eigenvalues of $\textbf{W}_m$ and $\textbf{W}_n$ in real-world bipartite networks including a small terrorists' social network, movielens network and netflix network. $\lambda_{\alpha}$ and $\tilde{\lambda_{\alpha}}$ are respectively the eigenvalues before and after coarse graining.}
\label{Table2}
\begin{tabular}{c c c c c c}
\hline
\hline
Network & $\alpha$ & $\lambda_{\alpha}$ ($\textbf{W}_m$) & $\tilde{\lambda_{\alpha}}$ ($\textbf{W}_m$) & $\lambda_{\alpha}$ ($\textbf{W}_n$) & $\tilde{\lambda_{\alpha}}$ ($\textbf{W}_n$)  \\
\hline
 &2 &0.8070&0.8059 &0.8070&0.8059\\
Terrorists &3&0.7259&0.7256&0.7259&0.7256\\
 &4&0.6013&0.5732&0.6013&0.5732\\
 \hline
 &2  & 0.4180& 0.4093 & 0.4180& 0.4093\\
Movielens &3  &0.2436 & 0.2305 &0.2436 & 0.2305\\
 & 4 & 0.2075 & 0.1890 & 0.2075 & 0.1890\\
 \hline
 &2  & 0.2575& 0.2535  & 0.2575& 0.2535\\
Netflix & 3  &0.2209 & 0.2168  &0.2209 & 0.2168\\
 & 4 & 0.2148 & 0.1971 & 0.2148 & 0.1971\\
\hline
\hline
\end{tabular}
\vspace*{0.0cm}
\end{center}
\end{table}

\begin{figure*}[htb]
    \centering
    \includegraphics[width=5.5cm]{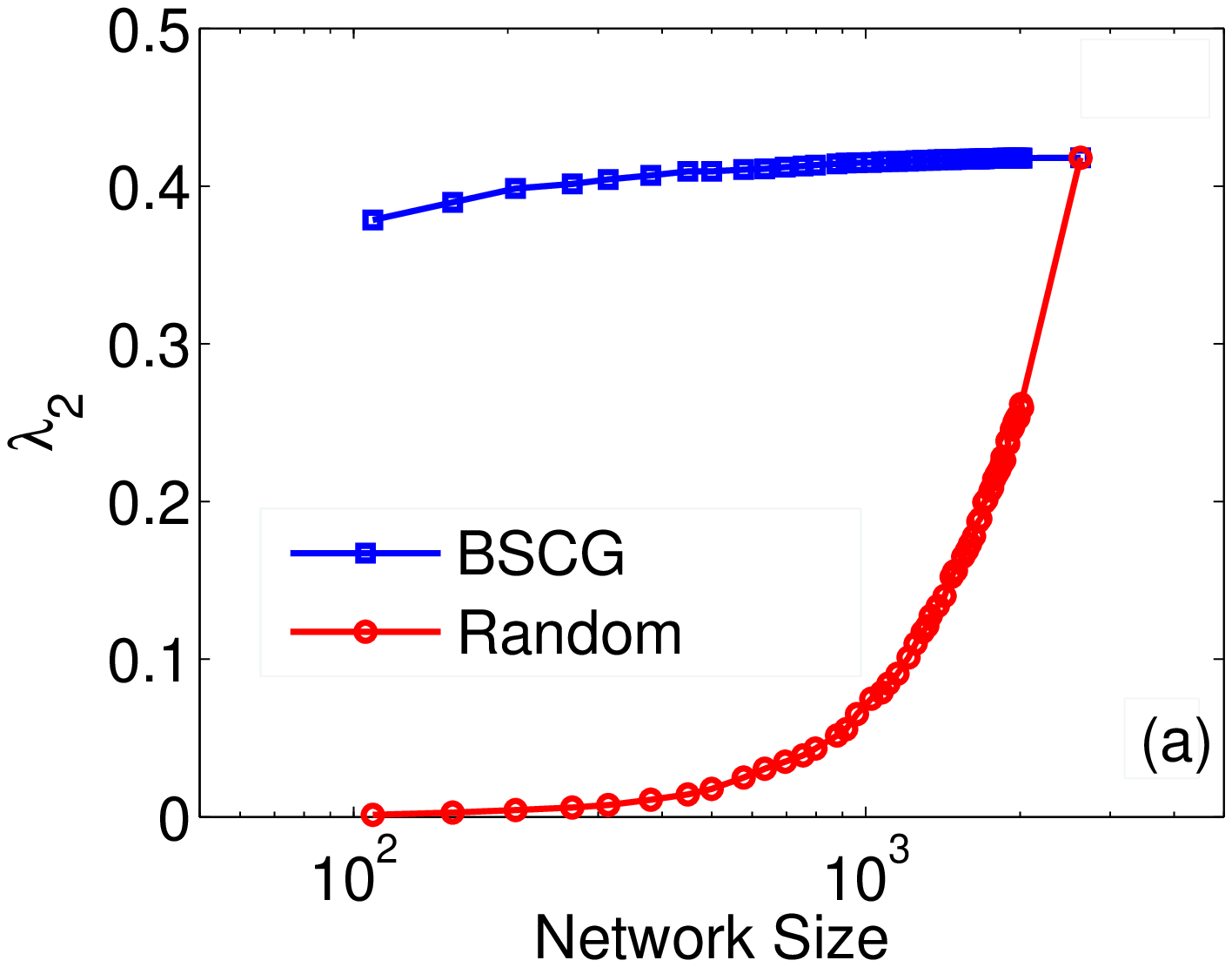}
    \includegraphics[width=5.5cm]{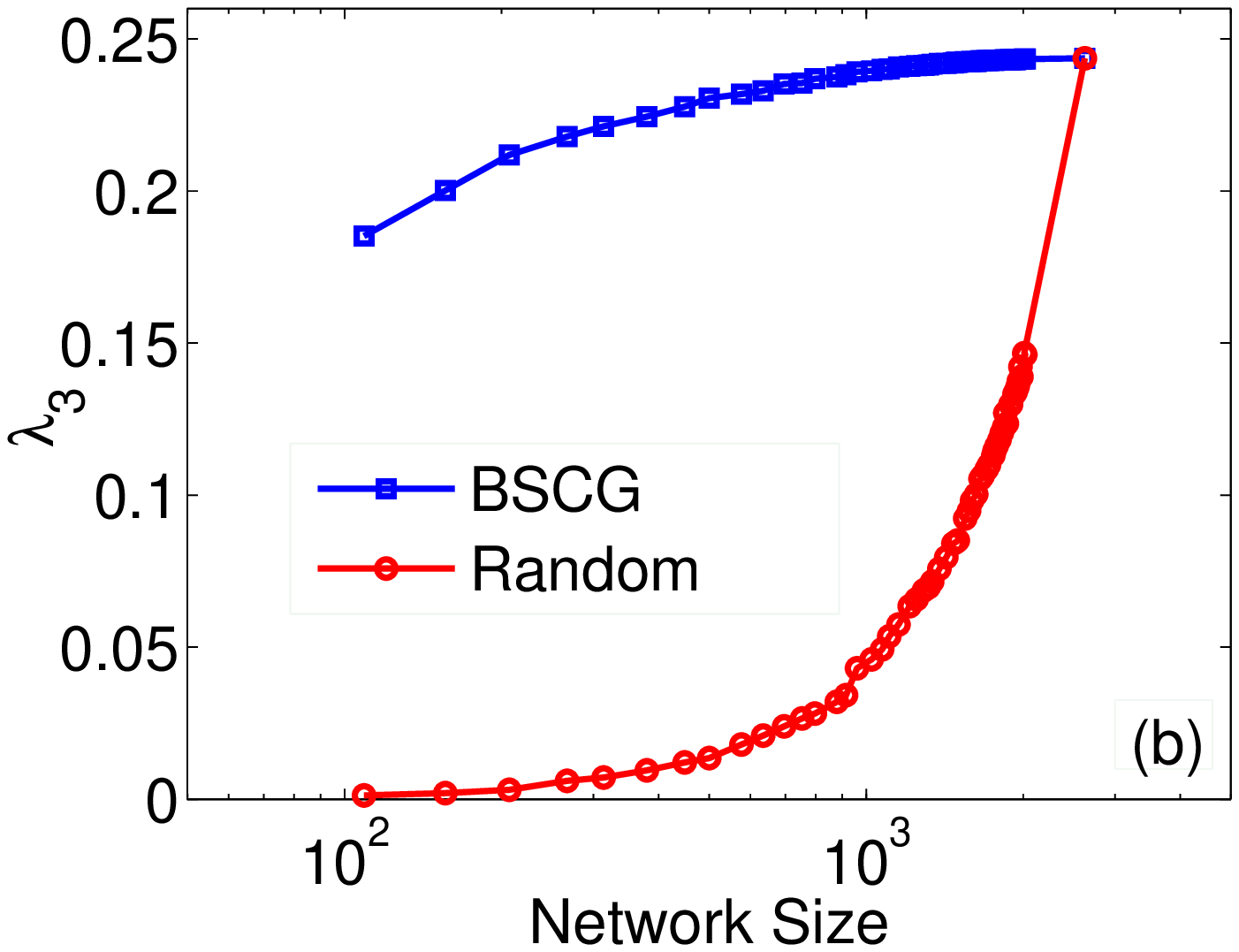}
    \includegraphics[width=5.5cm]{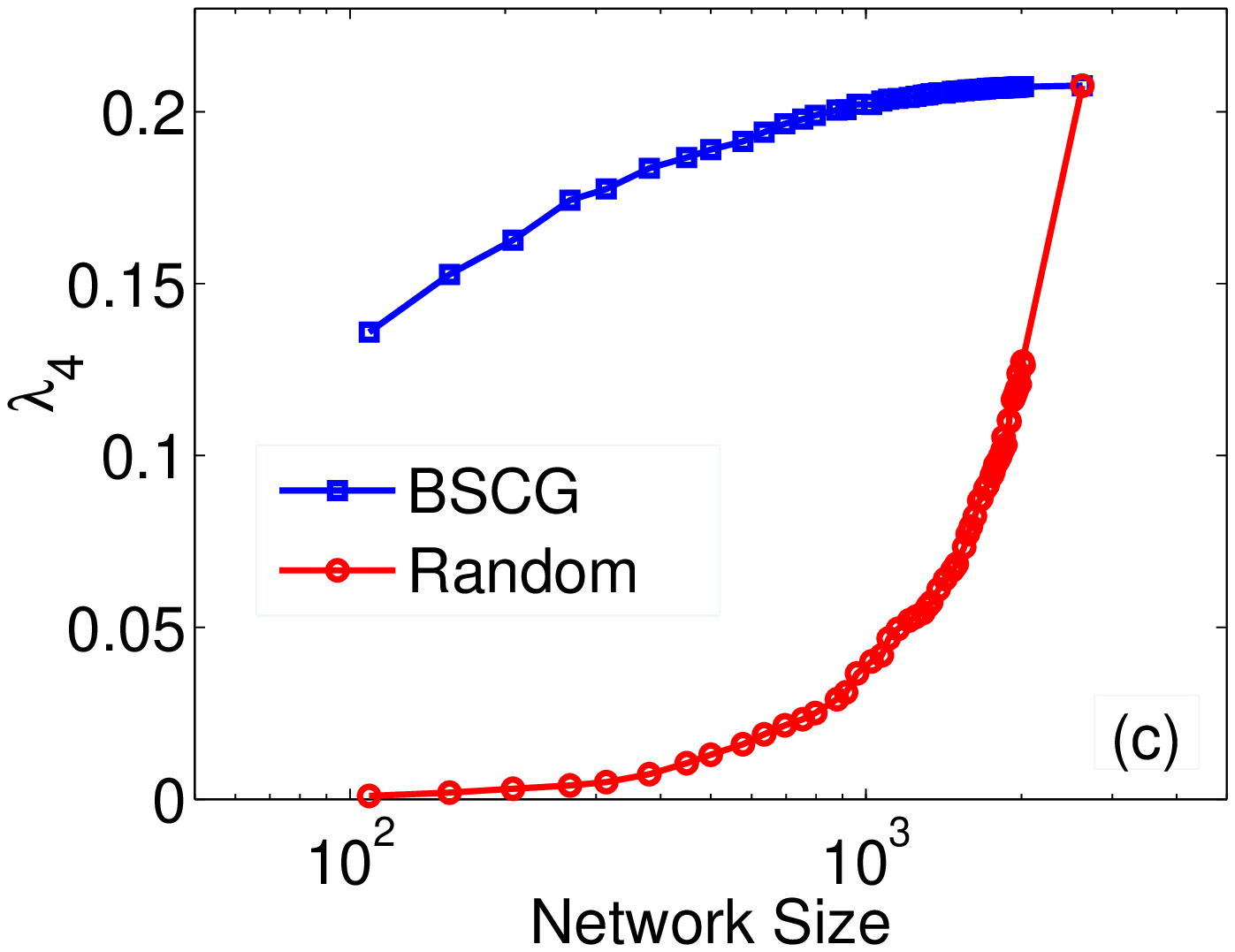}\\
    \includegraphics[width=5.5cm]{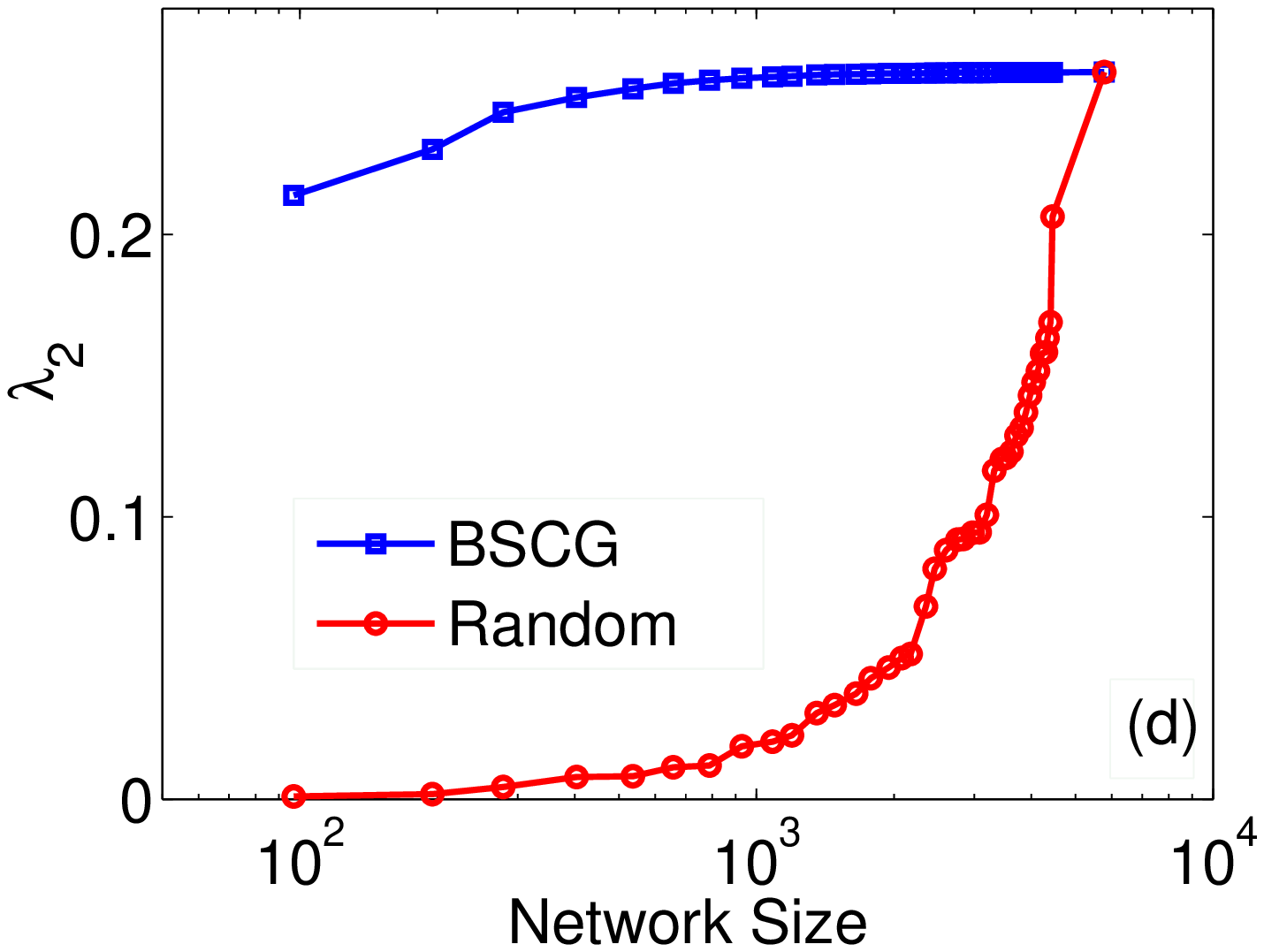}
    \includegraphics[width=5.5cm]{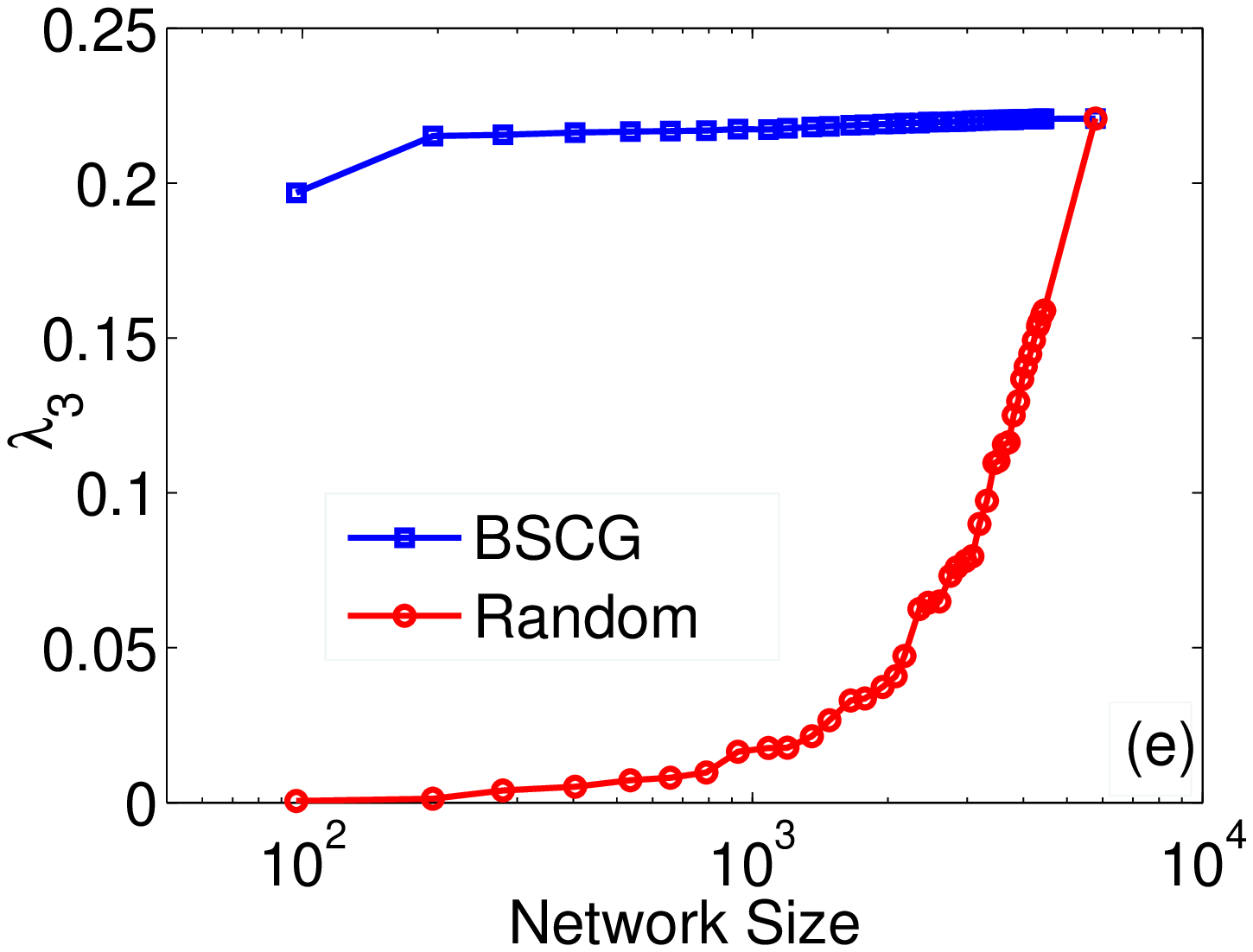}
    \includegraphics[width=5.5cm]{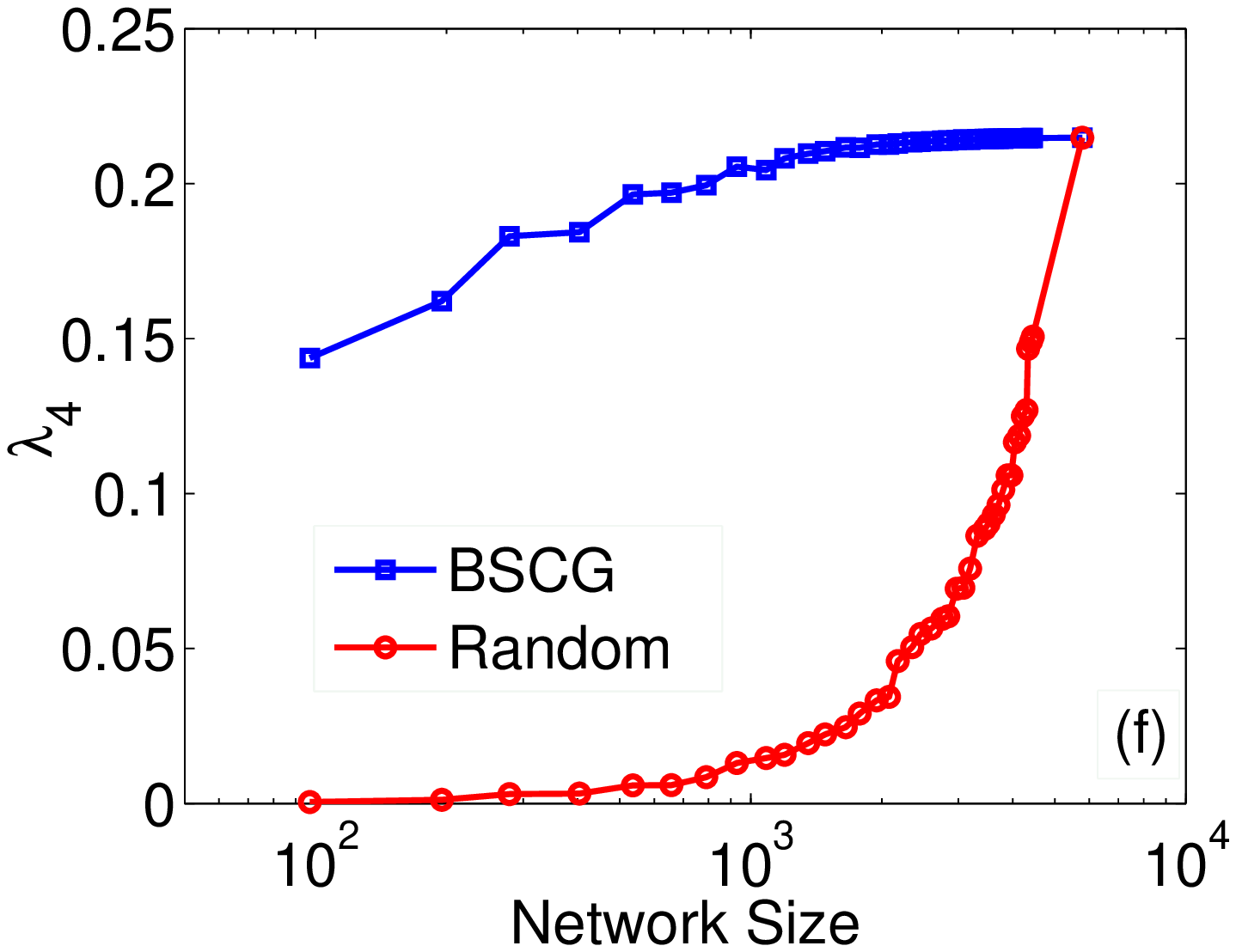}
  \caption{(Color online) The evolution of the three largest nontrivial eigenvalues $\lambda_2$, $\lambda_3$ and $\lambda_4$ as a
function of the size of the coarse-grained network. (a) - (c) The original network is movielens network. (d) - (f) The original network is netflix network. Red circles correspond to a random merging nodes, and the blue squares represent the BSCG Method. The results compared with random merging show the advantages of BSCG method when reducing the size of the bipartite network.}
\end{figure*}

In this subsection, we apply our method to some real-world networks. First, we apply the method to a social network of terrorists and the data was collected from 430 websites. The network was sampled from the data collected over a period from Oct. 1st, 1949 to May 1st, 2012, and it was based on the relationship between terrorists and their organizations. In this small social network, we focus on the giant component which are composed of 73 nodes in total, including 20 organizations and 53 people. The structure of the original network can be seen in Fig. 1, where the blue square accounts for the people and the red circle represents the organizations. The links between a person and an organization indicates that the person belongs to the organization. To coarse grain this network, we set $I = 5$ and obtain a reduced network with 23 nodes, which is shown in the bottom figure in Fig. 1. The three largest nontrivial eigenvalues before and after coarse graining are shown in Table \ref{Table2}. Clearly, all these eigenvalues are well preserved. Using our new method, the resulting network is also a bipartite network. Furthermore, we also try the method introduced in~\cite{Gfeller2007} on this real-world network, the resulting network is a binary one and the original two different kinds of nodes are indistinguishable.

As a further step, we apply our method to two online commercial networks: MovieLens and Netflix. The movielens network was sampled from the data collected over a seven-month period from September 19th, 1997 through April 22nd, 1998. The data consisted of 100,000 movie ratings from 943 users on 1,682 items. Each user sampled had rated at least 20 items. Users can vote for movies with five level ratings from 1 (i.e., worst) to 5 (i.e., best). Here we only consider the ratings higher than 2, so the data contains 82,520 user-object pairs. This sampled data is freely available at~\cite{website1}. The Netflix network was randomly sampled from the huge data set provided for the Netflix Prize. The original data is freely available at~\cite{website2}. It has 480,189 users, 17,770 items and 100,480,507 ratings. In the paper, we only consider a subset of this huge data set. The subset consists of 3,000 users, 2,779 movies, and 824,802 links. Similar to the MovieLens data, only the links with ratings no less than 3 are considered. After data filtering, there are 197,428 links left in the netflix network.

\begin{figure}[htb]
  \begin{minipage}[b]{0.5\columnwidth}
    \centering
    \includegraphics[width=\columnwidth]{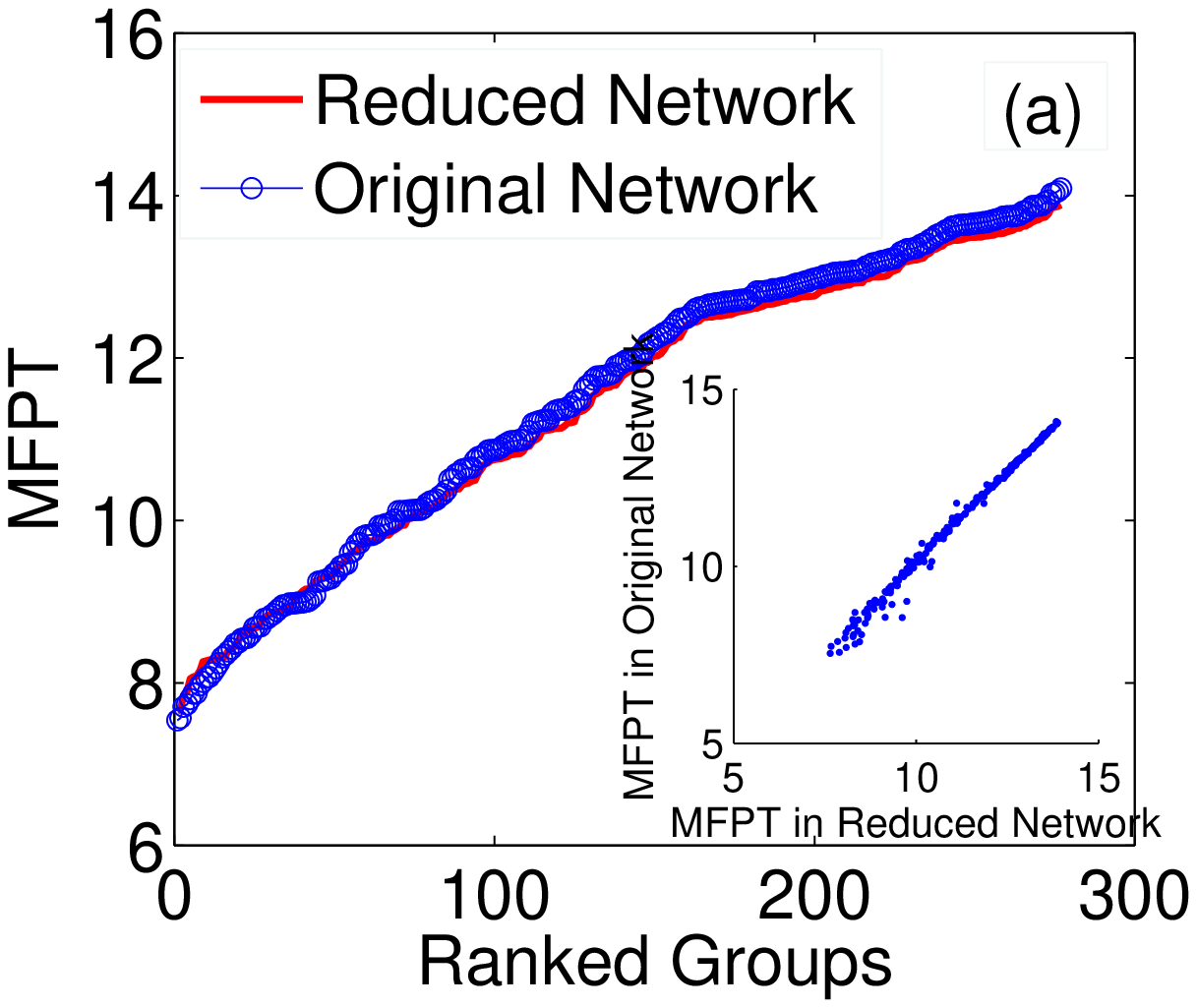}
  \end{minipage}%
    \begin{minipage}[b]{0.5\columnwidth}
    \centering
    \includegraphics[width=\columnwidth]{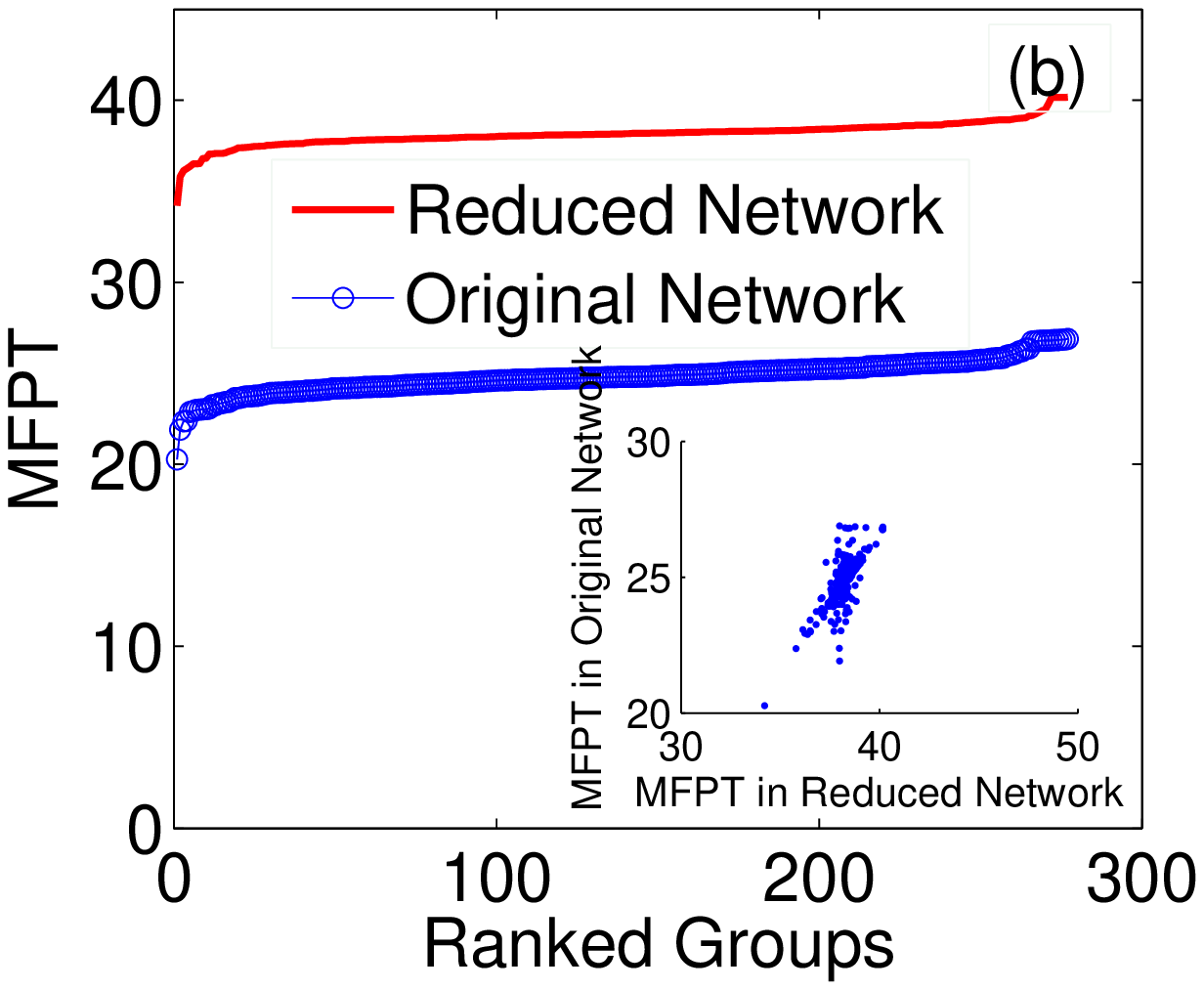}
  \end{minipage}
  \begin{minipage}[b]{0.5\columnwidth}
    \centering
    \includegraphics[width=\columnwidth]{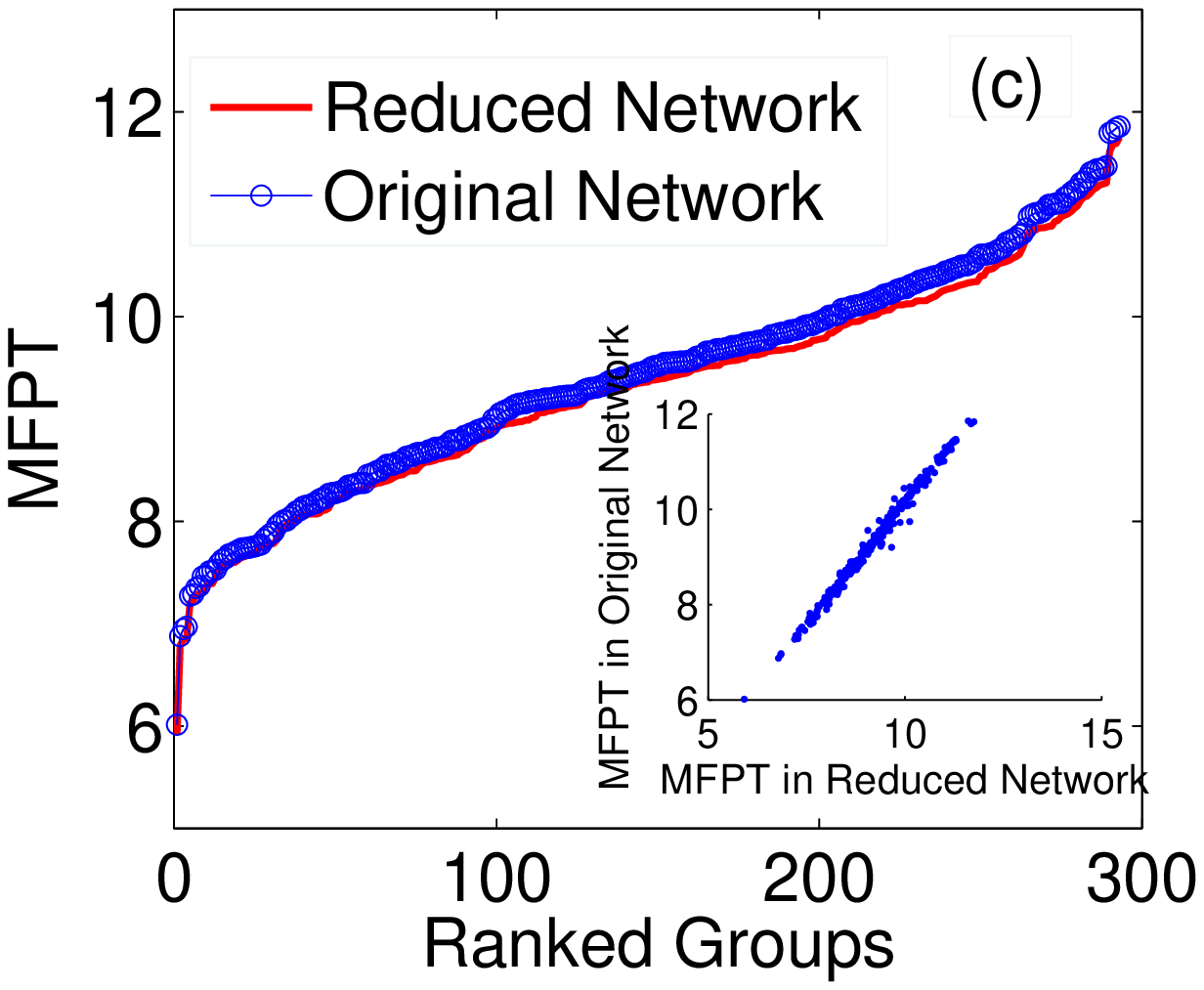}
  \end{minipage}%
      \begin{minipage}[b]{0.5\columnwidth}
    \centering
    \includegraphics[width=\columnwidth]{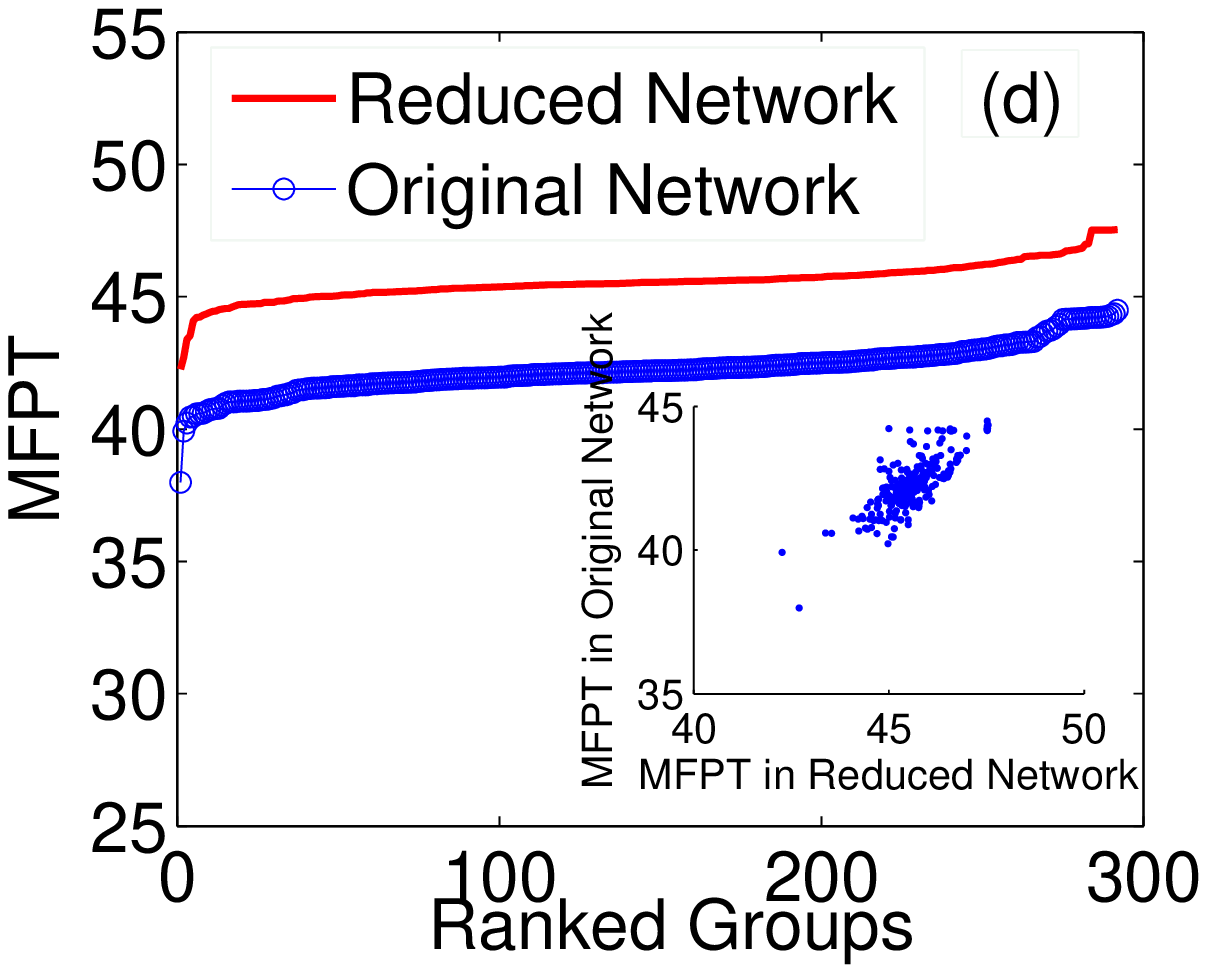}
  \end{minipage}
  \caption{(Color online) Comparison of the MFPT. The walker starts at each node in the top set and the sink $i$ is selected as the node with the strongest weight in the bottom set. The blue circles  represent the average MFPT ranked for each group in the original network. The MFPT of the corresponding nodes in the coarse-grained network is displayed with red lines. (a) The MovieLens network using BSCG method. (b) Nodes merged randomly in Movielens network. (c) The Netflix network using BSCG method. (d) Nodes merged randomly in Netflix network. Insets: Comparison of the exact MFPT between original and the reduced bipartite network. Slope $1$ represents the well preserved MFPT in the reduced network.}
\end{figure}

We first investigate how these three nontrivial eigenvalues evolve when the nodes in the networks are emerged. Fig. 2 shows the changes of these eigenvalues as a function of network size $N+M$. The red line corresponds to a random merging of the nodes into groups, and the blue curve shows the results when using the BSCG method in these eigenvalues $\lambda_2$, $\lambda_3$, and $\lambda_4$, i.e., nodes are grouped if their components in these eigenvectors according to $\lambda_2$, $\lambda_3$, and $\lambda_4$ are sufficient close to each other ($p_{\alpha}^i \approx p_{\alpha}^j$). The different values of network size $N+M$ correspond to different choices of the number
of intervals $I$ defined between the smallest and the largest
component in the eigenvectors. Generally speaking, a small $I$ yields a small network size. According to the results shown in Fig. 2, these three eigenvalues are well preserved even though the network size is significantly reduced. Actually, $I$ can be regarded as a parameter to determine the how accurate the eigenvalues are expected to be preserved, bigger $I$ can improve the precision of the method while resulting a bigger size of the reduced network.

In Fig. 2, it is also clearly shown that if nodes are merged randomly, the eigenvalues change dramatically . Consequently, the properties of random walk will be different from those in the original network. If the network are coarse grained according to the BSCG method, the nature of random walk in bipartite network could be effectively preserved. In order to keep eigenvalues almost unchanged, we set $I = 12$ in movielens network and get a reduced network with size $N+M = 500$, which is $20\%$ as big as the original network. The three largest eigenvalues in the reduced network can be seen in table \ref{Table2}. In netflix network, we set $I = 60$ and finally 657 nodes left which is about $10\%$ as big as the size of the original network. The table \ref{Table2} also shows the well preserved eigenvalues of $\textbf{W}_m$ and $\textbf{W}_n$ in netflix.

A more direct test of our method is to compare the mean first passage time (MFPT) from node $i$ to node $j$, which is denoted by $T_{ij}$ in the original and reduced networks. In the context of transport phenomena or navigation on a network, MFPT is an important characteristic of random walk. We label the nodes in the bipartite network from $1$ to $N'$ ($N'=N+M$) and consider the bipartite network as a binary one. In this way, all the cases for random walk in bipartite networks are included, i.e. the random walker can start from one type of nodes and finally arrive at either the same type or the other type of nodes. We consider multi-sink random walk problem and the MFPT can be exactly calculated by Eq.~\ref{equation3}.

In order to compare the MFPT between original and reduced network in movielens, we use the coarse grained network with $N'= 500$ obtained above. Specifically, we consider the walker starts at each node in the top set and define the node $i$ with the strongest node weight as the sink in the bottom set. In Fig. 3 (a), blue circles represent the MFPT from each node in the top set to nodes belonging to the group $i$ in the bottom set in the original network. The MFPT to the group $i$ in the bottom set in the reduced network is displayed with red lines. The exact overlap indicates that the MFPT is well preserved in the reduced network. The inset of Fig. 3 (a) shows the relationship between the MFPT of original network and that of reduced network. The result implies almost equal MFPT in both the original and the reduced network, given the same the source node and the sink. The slope of the curve is $0.996$ and the goodness of linear fit is $R^2 = 0.998$. However, the random coarse graining method significantly destroys the MFPT. As shown in Figure 3 (b), the MFPT between original network and reduced network differs from each other. From the inset of Fig. 3 (b), it is shown that there is no significant relationship between these two MFPT. We further test the MFPT in the Netflix network and its coarse gained networks from BSCG method and random method. Similar results are obtained (see Fig. 3(c) and (d)).

Besides computing the exact MFPT from Eq.~\ref{equation3}, we also use the numerical simulation of the random walk process to test the BSCG method. That is to say, we put a walker on each node in the bipartite network and it travels based on the stochastic matrices ($\textbf{U}$ and $\textbf{V}$). Similar results to Fig. 3 are obtained, namely the reduced network from BSCG effectively preserves the MFPT while the random coarse grained network significantly changes the MFPT. Finally, we remark that the results in Fig. 3 is consistent in different choices of sinks. No matter the walker starts and ends at nodes in the same or different set of nodes, the MFPT of the reduced network from BSCG method well overlaps with that of the original network. Taken together, the BSCG method is very robust in its performance.

\section{Conclusion}

One of the most difficult hurdles in the analysis of complex network is the huge size of the real-world systems. If the network has more than $10^5$ nodes, many algorithms are quite slow and sometimes even not doable. In order to solve this challenge, some coarse grain method based on complex networks are proposed, which mainly focus on the one-mode network in which only one type of nodes exist.

In this paper, we proposed a new coarse grain method for bipartite network with respect to random walk. After introducing two square stochastic matrices $\textbf{W}_m$ and $\textbf{W}_n$, we find that their three largest nontrivial eigenvalues can effectively represent the properties of random walk. To merge node with similar components of these eigenvectors corresponding to the eigenvalues, the reduced network with well preserved eigenvalues of stochastic matrix is obtained. Moreover, a straight test based on the mean first passage time is carried out in two real-world bipartite networks, and this property is well preserved in the reduced bipartite network. We believe that this method can be easily extend to preserve many other of spectral-determined dynamical properties in bipartite networks. Moreover, we have shown that for a bipartite network that the coarse graining provides a highly representative approximation of the initial network, giving rise to a way to circumvent the large size of complex networks for their analysis and visualization.

Finally, from a computational point of view, the
first eigenvectors are fast to calculate with the existing
optimized methods for sparse matrices, including the Lanczos and QL algorithms. Therefore our method can be easily utilized on large bipartite networks.

\section*{ACKNOWLEDGEMENTS}
This work is supported by the NSFC under grants No. 61174150 No. 70771011 and No. 60974084, NCET-09-0228, and fundamental research funds for the Central Universities of Beijing Normal University.

\end{document}